\documentclass[12 pt, ameircan, twoside, letterpaper]{article}
\usepackage[english]{babel}
\usepackage{color}
\usepackage{graphicx}
\usepackage{amssymb}

\title{Herbig-Haro objects around CG~30}

\author{\normalsize P. Kajdi\v{c} \\ \scriptsize Instituto de Geof\' isica, UNAM \\ \normalsize B. Reipurth \\ \scriptsize Institute for Astronomy, University of Hawaii, USA \\ \normalsize A. C. Raga \\ \scriptsize Instituto de Ciencias Nucleares, UNAM \\ \normalsize J. Walawender \scriptsize Institute for Astronomy, University of Hawaii, USA}

\begin{document}
\begin{displaymath}
\begin{array}{c}
\\
\\
\\
\\
\\
\\
\\
\\
\\
\\
\\
\\
\\
\\
\\
\\
\\
\\
\end{array}
\end{displaymath}
\begin{center}{\Large Rev. Mex. Astron. Astrophys., 2010, 46, 67}\end{center}
\maketitle
\abstract{In this work we study Herbig-Haro objects located in the
region around the head of the cometary globule CG~30. Two sets of
optical images are presented. The first set was obtained with the
3.5~m New Technology Telescope in 1995 in three emission lines:
H$\alpha$, S~II$\lambda\lambda$6731,6716~\AA\ and
[O~II]$\lambda$3729~\AA. The second set is an H$\alpha$ image of the
CG~30/31/38 complex obtained in 2006 with the 8~m Subaru telescope. A
proper motion study of the HH objects in the region was performed
using the H$\alpha$ images from both epochs.  Due to the high
resolution of our images we were able to, for the first time, resolve
the HH~120 object into ten knots and measure proper motions for some
of them.  We discover several new HH objects and
 a large bipolar jet, HH~950, emerging from
the head of CG~30.  We suggest that two previously known submillimeter
sources are the driving sources for the HH~120 and HH~950 flows.}

\section{Introduction}
\label{sec:Introduction}

Cometary globules (CGs) are compact interstellar molecular clouds
usually associated with H~II regions and OB stars.  A particularly
fine group is located in the Gum Nebula (Hawarden \& Brand 1976;
Zealey et~al. 1983; Reipurth 1983).  They show compact and
bright-rimmed heads and faint tails that extend from the heads and
point away from the nearby bright young photoionizing stars
(Fig.~1). Their sizes range between 0.1-0.8~pc, they exhibit high
densities, 10$^4$-10$^5$~cm$^{-3}$ and temperatures around 10 K. Their
typical masses range between 10-100~M$_\odot$. Cometary globules are
the sites of star formation (see Reipurth 1983). In some cases,
bipolar Herbig-Haro (HH) flows have been observed emerging from the
heads of cometary globules (Schwartz 1977; Reipurth 1983;
Pettersson 1984).

The largest known optical H~II region in the Galaxy is the Gum Nebula
(Gum, 1952). It is located in the Galactic plane ($l$=258$^\circ$,
$b$=-2$^\circ$) at a distance of about 450~pc, and is excited by
$\zeta$ Pup, $\gamma^2$ Vel, and in the past by the progenitor to the
Vela pulsar. Its linear diameter is 250~pc. With its apparent diameter
of 36$^\circ$\ it covers a large part of the southern sky in the
constellations Vela, Puppis, Pyxis, Canis Major and Carina.  One of
the most studied groups of CGs in the Gum Nebula is the CG~30/31/38
complex.

\subsection{CG~30 and HH~120}
In this paper we present two sets of very deep images of CG~30 (see
\S~2).  CG~30 has been studied by various authors at different
wavelengths.  It has been recognized as a region of star formation due
to its association with the Herbig-Haro object HH~120, first noted by
Westerlund (1963) and Reipurth (1981). 

The earliest spectroscopic measurements of HH~120 were made by
Pettersson (1984), who obtained spectra of two different parts of
HH~120, one being its brightest condensation (knot A). Its mean radial
velocity of -42~km~s$^{-1} \pm$12~km~s$^{-1}$ is consistent with the
value -39~km~s$^{-1}$ obtained by Schwartz \& Greene (2003). The
electron temperature T$_e$ and the electron density n$_e$ are found by
Pettersson (1984) to be 9100~K$\pm$400~K and 1700~cm$^{-3} \pm$600
~cm$^{-3}$, respectively. Also five IR sources associated with CG~30
were observed. It was argued that one of them, CG30-IRS4 (also known
as IRAS~08076-3556), is an energy source for HH~120 and its associated
reflection nebula.

Graham \& Heyer (1989) imaged HH~120 in the R (0.6~$\mu$m), J
(1.2~$\mu$m), H (1.6~$\mu$m) and K (2.2~$\mu$m) bands. Like Pettersson
(1984) they also detected a source CG30-IRS4 in H and K band images.

Scarrott et~al. (1990) performed optical polarization studies of a
nebula seen against the central dark region of CG~30 that contains
HH~120. It was confirmed that the so called CG~30 Nebula is 
predominantly a reflection nebula illuminated by the IR source CG30-IRS4.

Persi et~al. (1994) observed HH~120 and its energy source CG30-IRS4
in the near infrared, mm continuum, and an ammonia line. It was recognized
that the IRAS 08076-3556
source is a very young low mass Class~I source embedded
in a dense core of CG~30. The measurements of the 1.3~mm
luminosity and the bolometric luminosity show that their ratio is
actually closer to that of the Class~0 sources. The authors stated
that the strong 1.3~mm emission is probably due to a circumstellar
dust disk around the IRAS source.

Hodapp \& Ladd (1995) reported the discovery of eight
shocked objects in infrared images in the H$_2$ 1-0 S(1) emission
line. On the basis of their relative positions the authors concluded
that they form parts of two outflows propagating in approximately
perpendicular directions. None of the supposed flows could be directly
associated with the optical HH~120 object. 

Nielsen et~al. (1998) observed the cometary globules CG~30/31/38 in
$^{13}$CO, $^{12}$CO and H$_2$ lines. They proposed that CG~30 with a
mass of $\sim$10~M$_\odot$ is associated with another globule along
the line-of-sight having a mass close to 2~M$_\odot$ and moving with a
velocity of 4~kms$^{-1}$ with respect to the local standard of
rest. Also a dense molecular outflow associated with CG~30 was
detected in the $^{12}$CO~(J=1-0) line and its total mass was estimated to
be 0.28~M$_\odot$. They reported the maximum flow velocity to be
9~kms$^{-1}$ and its dynamical age to be 1.7$\times$10$^{4}$~years. This
flow is propagating in a direction perpendicular to the tail of
CG~30 and is originating from the position of the CG30-IRS4 source.

Bhatt (1999) discussed the role of magnetic fields in cometary
globules. The author found that in the case of the CG~30/31 complex, the light
from the stars in the region is polarized in the range from $\sim$0.1
to $\sim$4 per cent. The polarization vectors seem to be perpendicular
to the direction of the tails of CG~30 and 31, but, interestingly,
almost aligned with the direction of the molecular flow detected by
Nielsen et~al. (1998). Bhatt also claimed that if the polarization is caused
by dust grains aligned by the magnetic field, then the polarization
vectors must be parallel to the projected magnetic field in the
region. The tails of CG~30 and 31 are much shorter and
more diffuse than the tail of CG~22, where the magnetic field is found to
be parallel to the globules's tail. According to Bhatt, CG
morphology depends on the relative orientation of the cloud magnetic
field and the radius vector of the CG head from the central source of
radiation and the winds that produce the cometary tails. Long and
narrow tails are to be observed when the magnetic field is parallel to
the radius vector, while short and diffuse tails develop if those two
are perpendicular.

In their search for young Solar System analogues, Zinnecker et
al. (1999) observed four regions containing low luminosity sources
associated with extended reflection nebulosities, among which was also
CG~30. The images were obtained in near-IR (J, H and K)
broad-bands. In their images they observed the confines of the
$\sim$0.3~pc diameter dark globule with the very red CG30-IRS4 source
at the center and the bluer nebulosity just above it.

While the widely accepted distance of the Gum Nebula is 450~pc, Knude
et~al. (1999, 2000) and Nielsen et~al. (2000) suggested a distance
of 200-250~pc on the basis of color-magnitude ((V - I) - V)
diagrams and $uvby\beta$ photometry of the stars that appear to be
located in the CG~30/31/38 region. The distance to the globules
remains under debate, in this paper we adopt 450~pc.

In their survey of Bok Globules, Launhardt et~al. (2000) observed
CG~30 at submm wavelengths (850~$\mu$m). They discovered two sources
lying along a north-south direction and separated by a projected
distance of $\sim$9000~AU. The northern condensation was identified as
a possible driving source for the HH~120 jet (and with the CG30-IRS4
source) and the southern condensation was proposed to be associated
with the larger of the IR flows. The two sources were observed again
by Wu et~al. (2007) at 350~$\mu$m.

Nisini et~al. (2002) obtained spectra in the 1-2.5~$\mu$m wavelength
span. The most prominent features observed were [Fe~II] and H$_2$
lines. On the basis of H$_2$ emission they found that HH~120 consists of
multiple temperature components probably due to a slow, J type shock.

Kim et~al. (2005) studied the low-star mass formation in the
CG~30/31/38 complex. They obtained X-ray, optical and near-IR
photometry of the stars in the region and found 14 new pre-main
sequence (PMS) stars in addition to the 3 previously known stars in
the region. According to the authors, these PMS stars belong to two
groups: one group having ages of $\leq$5~Myr at $d =$200~pc, with
spectral classes K6 - M4, and the other group of F - G type stars with
ages of $<$100~Myr and $d \sim $2~kpc.  They conclude that there were
at least two episodes of star formation - ongoing star formation such
as in the head of the CG~30 cloud triggered by UV radiation from OB
stars, and a formation episode that may have been triggered $<$5~Myrs
ago by preexisting O stars, such as the progenitor of the Vela SNR and
$\zeta$~Pup.

Chen et~al. (2008a) observed CG~30 in the 3~mm dust continuum, in
N$_2$H$^+$~(1 - 0) emission line and at 3 - 8~$\mu$m wavelengths. The
authors detected two sub-cores inside CG~30. From the millimeter
continuum observations they derived the gas masses of the two
sub-cores to be 1.1~M$_\odot$ (northern sub-core) and 0.33~M$_\odot$
(southern sub-core). The authors classified the northern source as a
Class~I object and the southern source as a Class~0 protostar. The IR
observations revealed two perpendicular collimated bipolar jets that
coincide with the knots previously discovered by Hodapp \& Ladd
(1995). The N$_2$H$^+$~(1-0) emission maps revealed two cores
spatially associated with the mm continuum dust cores.

In a subsequent study, Chen at al. (2008b) studied CG~30 in the
$^{12}$CO(2-1) line and 1.3~mm dust continuum.  The $^{12}$CO(2-1)
observations showed the existence of two bipolar molecular flows
propagating in almost perpendicular directions. The authors suggested
that one of the flows is associated with the northern compact source
and the other one with the southern source. The northern flow exhibits
the projected length of 27,000~AU and the position angle
P.A. $\sim$128$^\circ$. The projected length of the southern flow is
20,000~AU and its direction of propagation (P.A.) is
$\sim$57$^\circ$. The velocities of the flows are low
($\lesssim$12~kms$^{-1}$).

In this work we study the kinematics of the HH objects in CG 30 by
measuring their proper motions. We combine our observations with the
existing data at IR and submm wavelengths in order to understand the
global outflow properties in the globule. Table~2 lists the
coordinates of all the HH objects that appear in our images. We have
discovered a large bipolar HH flow, which we catalogue as HH~950. This
flow is extending $\sim$12'2 in the northeast-southwest
direction, escaping from the interior of CG~30, and displaying a
complex structure with various working surfaces.

This paper is organized as follows. In Section~2 we present our data and
the reduction techniques applied. In Section~3 we present our methodology
and the results of this work. The latter are discussed in Section~4, and the
conclusions are given in Section~5.

\section{Observations}

\subsection{NTT images}

Observations with the ESO New Technology Telescope were carried
out on February 7, 1995 using the ESO Multi Mode Instrument (EMMI) and
three different narrow-band filters in the H$\alpha$,
[O~II]$\lambda$3729~\AA\ and [S~II]$\lambda\lambda$6731,6716~\AA\
emission lines (Figure~\ref{fig-colors}). The central wavelengths and
widths of these filters were 6568, 3725, 6728~\AA\ and 33, 69 and
75~\AA, respectively.  The total integration time for each of the
exposures was 30 minutes. The standard basic data reduction
procedures were carried out (bias subtraction and flat-fielding). The
average FWHM of the stars is 0.7~arcsec. The summary of all of the
exposures presented in this paper is given in Table~\ref{tab-obs}.

\subsection{Subaru images}

Observations with the Subaru Telescope were carried out on January 4,
2006 using the Subaru Prime Focus Camera (Suprime-Cam). This camera is
a mosaic of ten 2048 $\times$ 4096 pixel CCDs, which are located at
the prime focus of the 8m Subaru Telescope. Narrowband filter N-A-L659
with central wavelength $\lambda_0$6600~\AA\ and FWHM 100~\AA\ was
used, thus including the H$\alpha$ line. Five exposures of CG~30 with individual exposure times of 360 seconds were
obtained.  Data reduction was performed using IRAF. The standard basic
data reduction procedures were carried out (i.e. bias and dark
subtraction, flat-fielding, and distortion correction). The images
were then stacked using the IRAF MSCRED, MSCSETWCS, MSCZERO and
GREGISTER packages. Figure~1 shows a region of the CG~30/31/38 complex
and is a small subimage of this mosaic.  This is the deepest image of CG~30
obtained so far. Its angular resolution is not as good as in the
NTT images (the average FWHM of the stars is $1''3$), probably
due to the fact that for the Subaru telescope CG~30 was
observed at high airmass.

\begin{table}
\caption{Observing log}
\begin{center}
\begin{tabular}{cccccc}
\hline
\hline
Date & Total & Telescope & Instrument & Central & $\Delta\lambda$ (\AA)\\
 & exposure time  &  &  & wavelength (\AA) & \\
\hline
\hline
February 7 1995 & 30 min & NTT & EMMI & 6568 & 33 \\
February 7 1995 & 30 min & NTT & EMMI & 3725 & 69 \\
February 7 1995 & 30 min & NTT & EMMI & 6728 & 75 \\
January 4 2006 & 30 min & Subaru & Suprime-Cam & 6600 & 100\\
\hline
\end{tabular}
\end{center}
\label{tab-obs}
\end{table}

\section{Methods and results}

\subsection{HH flows and objects}

The first HH object discovered in CG~30 was HH~120.  Here we present
the so far deepest and most detailed images of this HH flow.
Pettersson (1984) resolved the object into several knots, and we
retain the nomenclature A and B for the two main knots in HH 120.  We
resolve numerous knots, which we label A to J (see
Figure~\ref{fig-sub}). Knot A is the brightest and has been observed
previously at infrared wavelengths by Schwartz \& Greene (2003). These
authors obtained an image of HH~120 in the 1-0~S(1) line of H$_2$. The
image was used to locate the slit positions for spectroscopic
measurements. It shows two knots the authors call A and B that are
$\sim$4''6 apart.  The IR knot A coincides with the A knot in our
H$\alpha$ images (Figure~\ref{fig-sub}). The IR knot B does not
coincide with any of the knots in the H$\alpha$ line, but it lies just
northward of our knots C and D. The knots C and D lie between the
optical knots A and B and the knot E lies about 7''3 east of the knot
A. About 3''\ northwest of the knot E there is a small knot F. Four
faint knots, G, H, I, and J lie very close to the knot A just west of
it. Finally, in the Figure~\ref{figure-combine} can be seen the bright knot K, which lies 11''6 to the north of the knot A.

In the NTT H$\alpha$ image all of these knots appear
except for the knot F, which blends in the
background noise and the knot K, which lies outside the region showed in this image. The lower image in Figure~\ref{fig-sub} shows
a contour plot of HH~120. Only the two major condensations A and E are well
resolved by the contours.

In order to have a better overview of the HH objects in the region, we combined the three NTT images into a single image (see Figure~\ref{figure-combine}). The H$\alpha$ image was weighted with a factor $0.25$. The image in the Figure~\ref{figure-combine} was obtained by averaging the weighted H$\alpha$ image with original [S~II] and [O~II] images. 

Several new HH objects, here labelled HH~948, 949, and 950, appear in
our images. Some of them had already been observed by Hodapp \& Ladd
(1995) in the near-infrared.  In their work the authors labeled
detected IR flows with numbers from 1 to 8. We adopt the same
notation when we refer to them. Among the IR flows the following
appear in our [S~II] image: HH~948 (IR knot 2) consists of three knots that are located between 38 and 43''\ north of
the HH~120 knot A. IR knot 6 is located 52''\ southeast of the HH~120 knot A. HH~949 is composed of eight knots of which six (A to F) were previously known as IR knot 7, one of them (H) as IR knot 8 and one (G) as IR knot 5 (see Figure~\ref{hh949}). The knots A to F and H lie between 52''\ and 72''\ northeast of HH~120. HH~949G knot (IR knot 5) lies about 39''\ east of the HH~120 knot A.

None of these features appear in the [O~II] image,
while most of them (HH~949 and HH~948) do appear in our H$\alpha$ images, although they are very faint.

We discovered a new Herbig Haro flow HH~950. It is seen in all of
our images. This is a major bipolar flow that is emerging from the
head of the CG~30. It extends approximately 10'\ in the
northeast-southwest direction. It is composed of two lobes - the
northeastern and the southwestern lobe. The width of the flow at the
position of the HH~950 knot E (see Figure~\ref{figure-combine}) is
1', which yields a length to width ratio of 10  and an opening
angle of $\sim$5.7$^\circ$. The only feature observed in the
northeastern lobe is its rim. The 'forward' rim is brighter than the
'back' rim. They both appear in the form of long, thin filaments that
are best seen in the H$\alpha$ images. The southwestern lobe shows a
more complex structure. The filaments are typically about $\sim$ 2''\
wide and more than 40''\ long. The flow also contains six knots: the
two brightest knots are C and E which lie 4'\ and 2'{}4 from
HH~120 along the southwestern lobe.  Four smaller knots are visible
inside the same flow. They are knots A, B, D and F and lie at an
angular distance of 4'{}3, 3'{}9, 3'{}1 and 1'{}5 from
HH~120. Surprisingly they appear brightest in the [O~II] image.  Both
of the lobes are slightly curved in the direction towards the center
of the Gum Nebula. The angle of curvature of the southwest lobe is
larger and measures about 15$^\circ$.

\subsection{Kinematics}

In order to study the kinematics of the outflows in the CG~30 complex,
we derived proper motions of HH objects that appear in our images. The
two images on which we based our measurements are the H$\alpha$ NTT
image and the Subaru image. The time elapsed between the images is
10.91 years (3985 days).  Both H$\alpha$ images were registered so
they had the same pixel size, distortion and orientation. This
was done by using the IRAF GEOMAP and GREGISTER tasks. In order to
calculate the proper motions of chosen features, we used a 2D cross
correlation technique. The results are presented in
Table~\ref{tab-pm}.

The quantities of the columns in Table~\ref{tab-pm} are: (1) labels of
features, (2) proper motions in arcsec per century, (3) velocities in
kms$^{-1}$ assuming a distance of 450~pc, and (4) position angles of
proper motion vectors in degrees.

The values of proper motions of the knots that form parts of the
HH~950 flow are between 1.4 and 4''2/century and are oriented
towards the southwest (position angles between 240-253$^\circ$),
while the HH~120 knots propagate in a northwest direction (position
angles between 309-332$^\circ$) and have proper motion values
$1''3$ and $2''1$/century. At an assumed distance of 450~pc, the
projected velocities of the proper motions for the HH~950 knots are between 31
and 107~kms$^{-1}$ and for the knots associated with HH~120 they are
26 and 45~kms$^{-1}$.

We divided the HH~120 knots in two groups - knot E as one and all the
other knots as the second group. We got two somewhat different proper
motion vectors - the knot E seems to be propagating more westward than
the other knots. From Figure~\ref{fig-pm} we see that the proper
motion vector of the second group of knots of the HH~120 flow points
directly towards HH~948. 

Among the IR flows we obtained proper motion vectors for HH~949
(IR knots 5, 7 and 8).  The calculated direction of motion is
perpendicular to the axis on which six IR knots lie. The previous
surveys have not revealed any young stellar object that could be a
driving source for this HH object. We believe that this proper motion
vector is affected by changes in the structure of the HH flow between
the two epochs.
 
The proper motion vector of HH~948 is of particular interest. The
value of the proper motion ($1''4$/century) and the direction of the
proper motion vector (308$^\circ$) are very similar to the proper
motion vector of the HH~120 knots A, B, C and D ($1''3$/century and
309$^\circ$). Both vectors lie along the line connecting these HH~120
knots and HH~948, and so may be parts of the same HH flow.

\section{Discussion}

Given the structural complexity of HH 120 together with our proper
motion results, we believe that the HH~120 flow consists of at least
two outflows, and possibly more. Knots A, I, J and K appear to form
one flow. Knot G shows a long streamer to the west. Knots C, A and H
could form a bipolar flow, and knot E possibly forms a separate
flow. Its proper motion vector points more westward than the proper
motion vector for the rest of the HH~120 knots.

Launhardt et~al. (2000) discovered two sources in their survey in
their submillimeter 850~$\mu$m survey, lying in a north-south direction,
which we call here CG30~SMM-N and SMM-S. The SMM-N source is probably
a driving source for HH~120. Since it seems that HH~120 consists
of at least two flows, the SMM-N source could actually be a binary or
a multiple system.

The proper motion vector calculated for the group of knots that includes the knots C, A and H, points directly
towards HH~948. The directions and the absolute values of the
proper motions of these knots coincide well with the values of the
proper motion of HH~948. We conclude that HH~948 is
related to the flow to which knots C, A and H belong. 
The fact that there are other HH objects
present around HH~120 gives further support to the assumption that
there must be several outflows coming from SMM-N.

In order to put the IR and optical data in context, we combined both
images into a single image (see Figure~\ref{fig-krizi}). The IR and
optical HH knots are displayed there simultaneously. The two crosses
that appear in the image mark the positions of the two submm sources
discovered by Launhardt et~al. (2000).  It can be seen that SMM-S lies
exactly on the
axis defined by the IR knots 1, 3, 4, 5, 7 and 8. 
This makes it a perfect candidate for a driving source of this IR flow
(as already proposed by Launhardt et~al. 2000). However, is it also a
driving source of the large HH~950 flow?

When examining the positions of these IR knots and the HH~950 knots at
optical wavelengths we see that they do not lie on the same
axis. Especially intriguing are the positions of the HH~950 knots B, C
and F. The knots B and C are located in parallel positions along the
southwestern lobe.  The axis formed by the knot F, six IR knots and
SMM-S also does not coincide with the one
connecting SMM-S to any other feature that belong
to the HH~950 flow.  We compared the HH~950 flow with the HH~184
(Devine et~al. 1999). This object exhibits a very complex and somewhat
similar structure. It also shows HH knots that are located parallel
along the flow's axis. The authors of this work argue that the axis of
this flow changes because its source is a primary star in a binary
system with eccentric orbit. Whenever a secondary star (which emits a
smaller jet) approaches the primary the flow's axis changes
rapidly. Because of this some of the knots produced in that moment
appear to lie off-axis. When the secondary is further away from the
primary (which is the majority of time, due to the high eccentricity
of its orbit), the flow's axis returns to its ``normal'' position.

If the same explanation applies for the HH~950 jet, then SMM-S also
must be a binary system. However our arguments are based only on the
structure of the HH~950 jet. We do not see a secondary jet that would
confirm the binary nature of the HH~950 source.

There is also a question of the curvature of the HH~950 jet. It is
slightly curved in the direction of the center of the Gum Nebula. 
This could be another indication that the driving source is a binary.

\section{Conclusions}

In this work we have studied Herbig-Haro objects associated with
CG~30. We find that most of the HH objects belong to two flows: HH~120
and HH~950. The proper motions of the knots in HH~120 suggest that
this object is actually composed of at least two outflows.  The
candidate for the driving source of HH~120 is the northern submm
source CG30~SMM-N discoverd by Launhardt et~al. (2000), which must
therefore be a binary or a multiple system.

The southern submm source CG30~SMM-S is located exactly on the axis defined by
six IR objects discovered by Hodapp \& Ladd (1995). This leaves no
doubt that these objects really do form one flow and that this submm
source is its driving source (as was already suggested by Launhardt et
al. 2000).

We propose that this flow forms a part of the large HH~950 flow that
appears in our optical images. HH~950 consists of two lobes -
while its northeastern lobe is almost featureless (the only visible
feature is its bright rim), the southwestern lobe shows a complex
structure made of fine filaments and knots. Interesting is the fact that the HH
knots in this lobe do not all lie on the same axis. It seems that the axis
swiftly changed toward the south at the time the knots B and C were
produced.  We argue that CG30~SMM-S could be a binary system. At
least one member of the system is a young star emitting the HH~950
jet. The interaction of both objects causes the change of direction of
the HH~950 axis.

{\bf Acknowledgments}\\
We are grateful to an anonymous referee for a very careful and  
helpful report. We thank Ralf Launhardt for providing the accurate
positions of the submillimeter sources and Klaus Hodapp for providing
us with the original infrared image of the CG~30.  Primo\v{z}
Kajdi\v{c} acknowledges the Direcci\'on General de Estudios de
Posgrado of the UNAM for a scholarship supporting his graduate
studies.  This study has been supported by the NSF through grant
AST0407005.  This material is based upon work supported by the
National Aeronautics and Space Administration through the NASA
Astrobiology Institute under Cooperative Agreement No. NNA04CC08A
issued through the Office of Space Science.

\begin{table}\centering
\caption{Coordinates of the observed HH objects in the region around CG~30}
\begin{tabular}{ccccc}
      \hline
      \hline
      Feature name & R.A. (hh:mm:ss.s) & Dec. (dd:mm:ss) \\
      \hline
      \hline
\multicolumn{3}{c}{HH~948}\\
A & 08:09:30.6 & -36:04:25\\
B & 08:09:30.6 & -36:04:19\\
C & 08:09:31.8 & -36:04:16\\
\hline
\multicolumn{3}{c}{HH~949}\\
A & 08:09:36.8 & -36:04:29\\
B & 08:09:37.2 & -36:04:27\\
C & 08:09:37.4 & -36:04:26\\
D & 08:09:36.8 & -36:04:25\\
E & 08:09:37.2 & -36:04:25\\
F & 08:09:36.5 & -36:04:35\\
G & 08:09:35.7 & -36:04:50\\
H & 08:09:37.7 & -36:04:20\\
\hline
\multicolumn{3}{c}{HH~120}\\
A & 08:09:32.5 & -36:04:55 \\
B & 08:09:32.9 & -36:04:56 \\
C & 08:09:32.6 & -36:04:56 \\
D & 08:09:32.8 & -36:04:56 \\
E & 08:09:33.1 & -36:04:56 \\
F & 08:09:33.0 & -36:04:55 \\
G & 08:09:32.3 & -36:04:58 \\
H & 08:09:32.3 & -36:04:55 \\
I & 08:09:32.5 & -36:04:53 \\
J & 08:09:32.5 & -36:04:51 \\
K & 08:09:32.4 & -36:04:44\\
\hline
\multicolumn{3}{c}{HH~950}\\
A & 08:09:14.6 & -36:07:15 \\
B & 08:09:16.5 & -36:07:12 \\
C & 08:09:17.8 & -36:07:27 \\
D & 08:09:22.1 & -36:07:07 \\
E & 08:09:24.6 & -36:06:40 \\
F & 08:09:28.6 & -36:06:10 \\
\hline
\multicolumn{3}{c}{The submm sources (from Wu et~al. (2007)).}\\
North & 08:09:32.87 & -36:04:56.3 \\
South & 08:09:32.55 & -36:05:16.6 \\
\hline
\end{tabular}
\label{tab-feathh120}
\end{table}

\begin{table}
\caption{Proper motions: velocity and angle}
\begin{center}
\begin{tabular}{cccc}
\hline
\hline
Feature & Proper motion ( ''/century) & kms$^{-1}$ & P.A. (degrees)\\ 
\hline
\hline
HH~120A, B, C, D  & 1.3 &26 &  332\\
HH~120E  & 2.1 & 45& 309\\
HH~948B & 1.4 &  31&308\\
HH~950A & 3.8 &  82&253\\
HH~950C & 4.2 &  89&240\\
HH~950D & 1.8 &  38&250\\
HH~950E & 5.0 &  107&244\\
\hline
\end{tabular}
\end{center}
\label{tab-pm}
\end{table}

\begin{figure}
\centering
\includegraphics[width=1.0\textwidth]{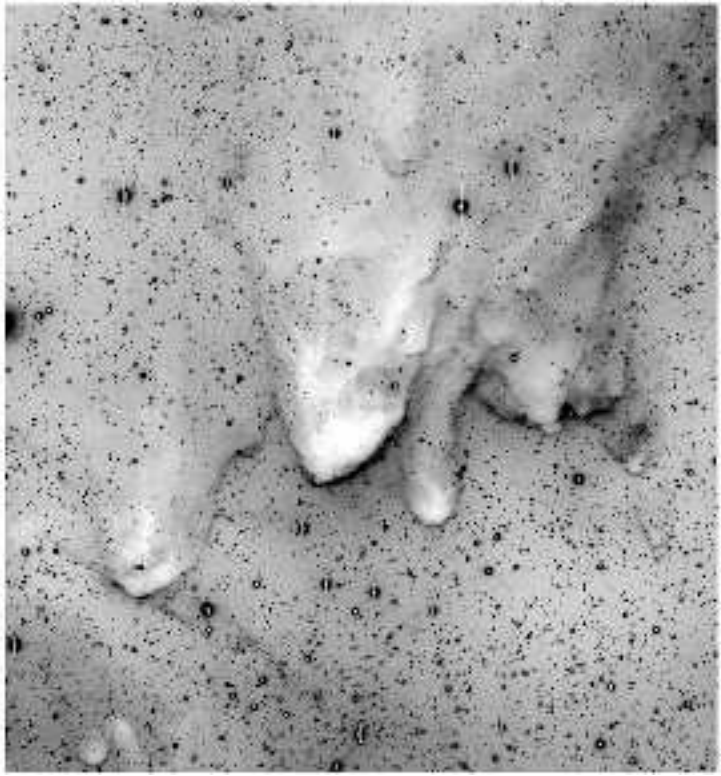}
\caption{The H$\alpha$ image of the CG~30/31/38 complex obtained with
the Subaru telescope and Subaru Prime Focus Camera on January 4, 2006,
composed of five six-minute exposures. This is the deepest image of
the complex ever obtained. The angular size of the image is 21'6$\times$23'5. North is up and east is left.}
\label{fig-cg30}
\end{figure}

\begin{figure}
\centering
\begin{tabular}{cc}
\includegraphics[width=0.5\textwidth]{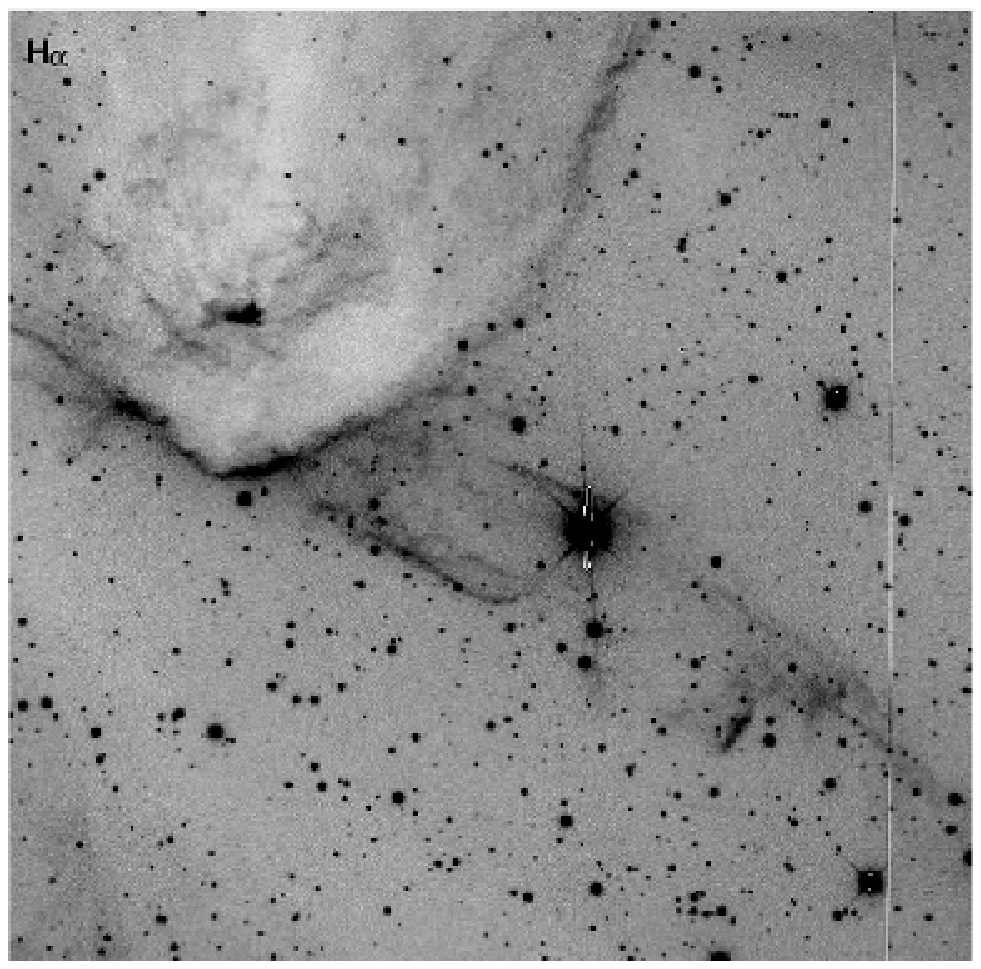}&
\includegraphics[width=0.5\textwidth]{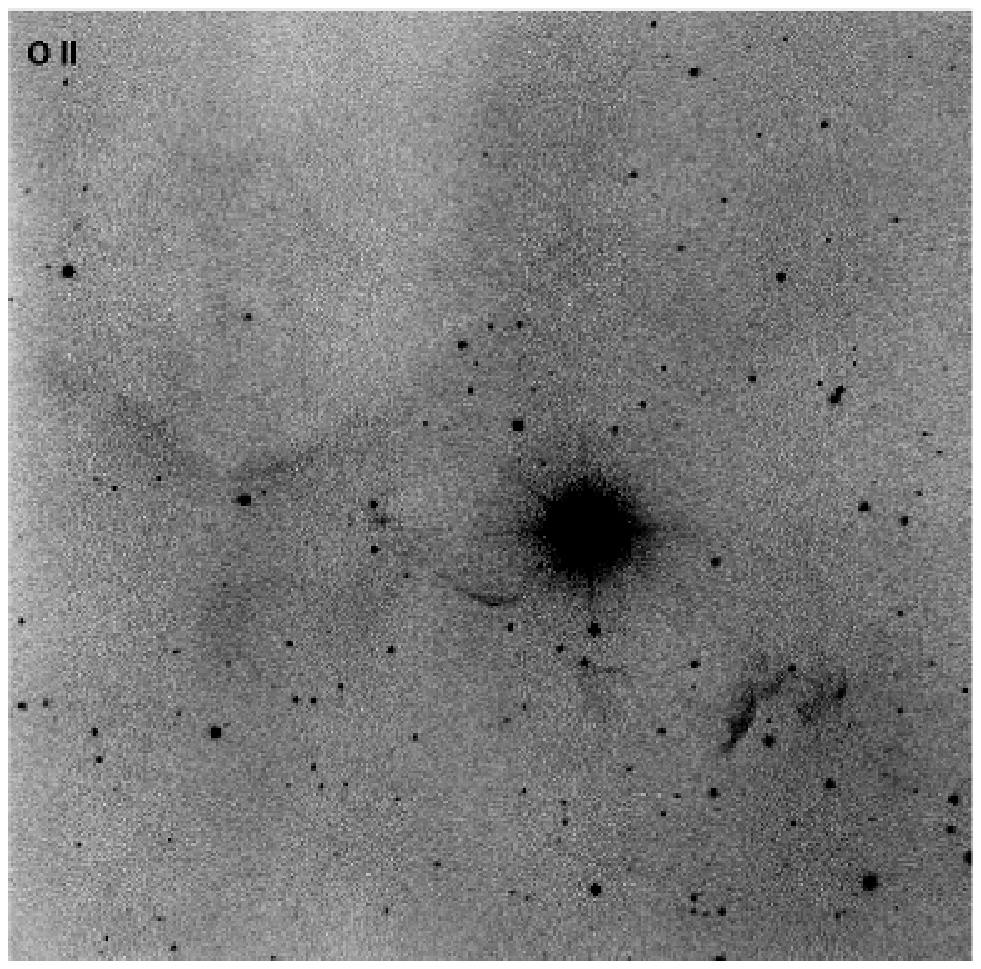}\\
\includegraphics[width=0.5\textwidth]{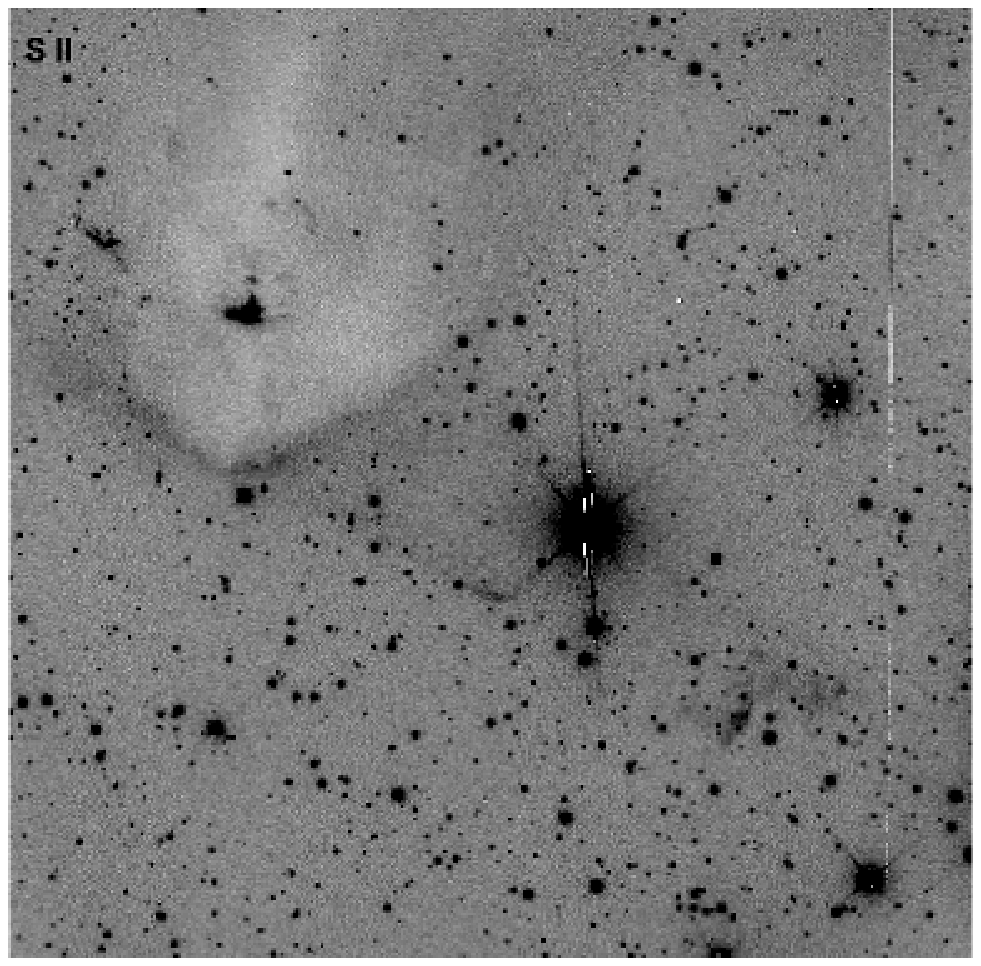}&
\end{tabular}
\caption{The three images were obtained with ESO NTT telescope and ESO Multi
Mode Instrument, using three narrow band filters: H$\alpha$ (top),
[O~II]~$\lambda$3729~\AA\ (medium) and [S~II]~$\lambda$6730+6716~\AA\ (bottom). Each image is a 30 minute exposure.  The angular size of the images is 5'8$\times$5'7. North is up an east is left.}
\label{fig-colors}
\end{figure}

\begin{figure}
\centering
\begin{tabular}{c}
\includegraphics[width=1.0\textwidth]{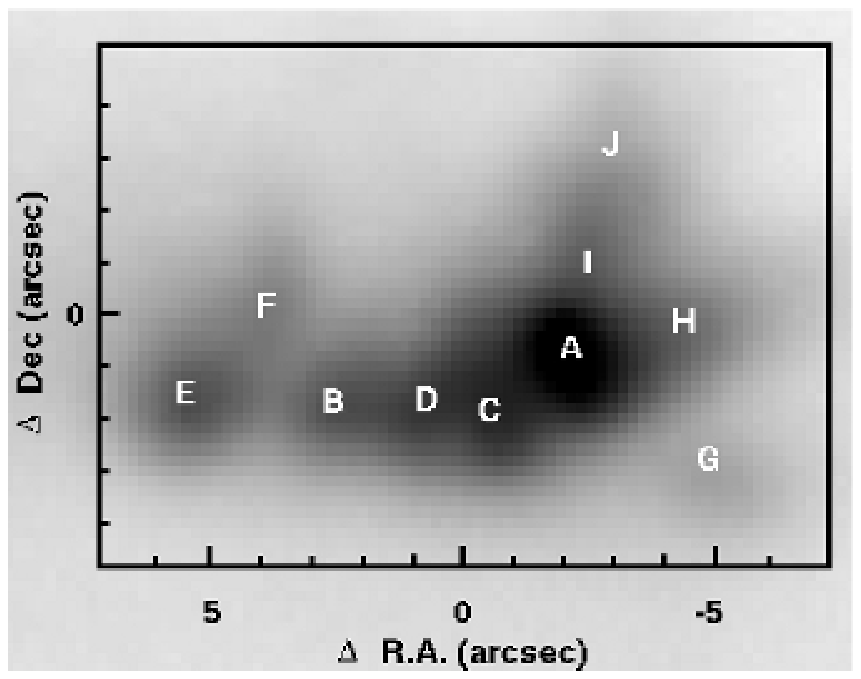}\\
\includegraphics[width=1.0\textwidth]{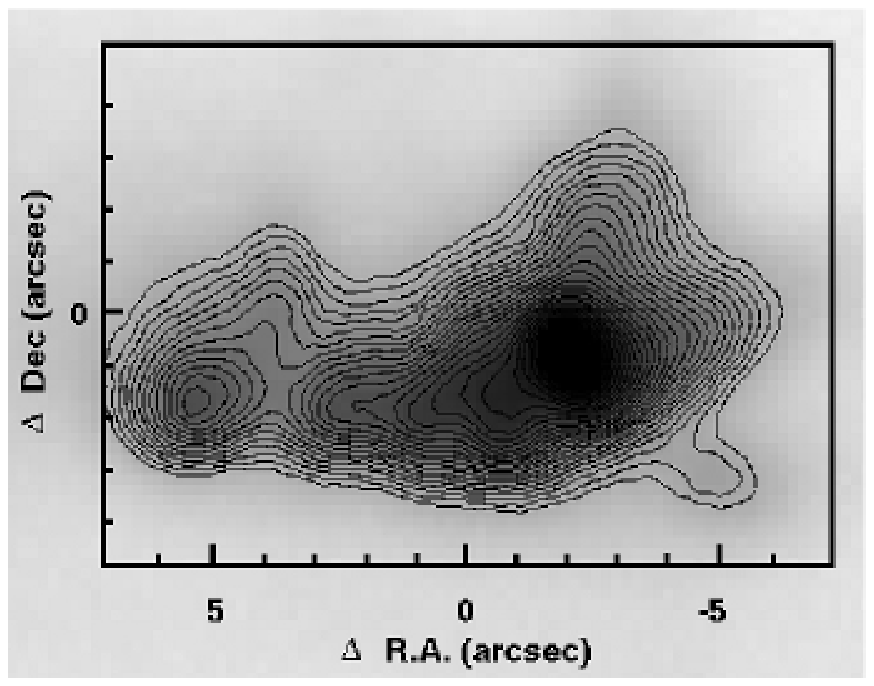}
\end{tabular}
\caption{The detailed Subaru H$\alpha$ images of HH~120. In addition
to the knot A which had already been observed, we resolve ten other
knots: B, C, D, E, F, G, H, I, J and K.  Bottom: Logarithmically
scaled countours are plotted on top of the HH~120 image. Only knots A,
E and G are resolved by the contours.}
\label{fig-sub}
\end{figure}

\begin{figure}
\centering
\includegraphics[width=1.0\textwidth]{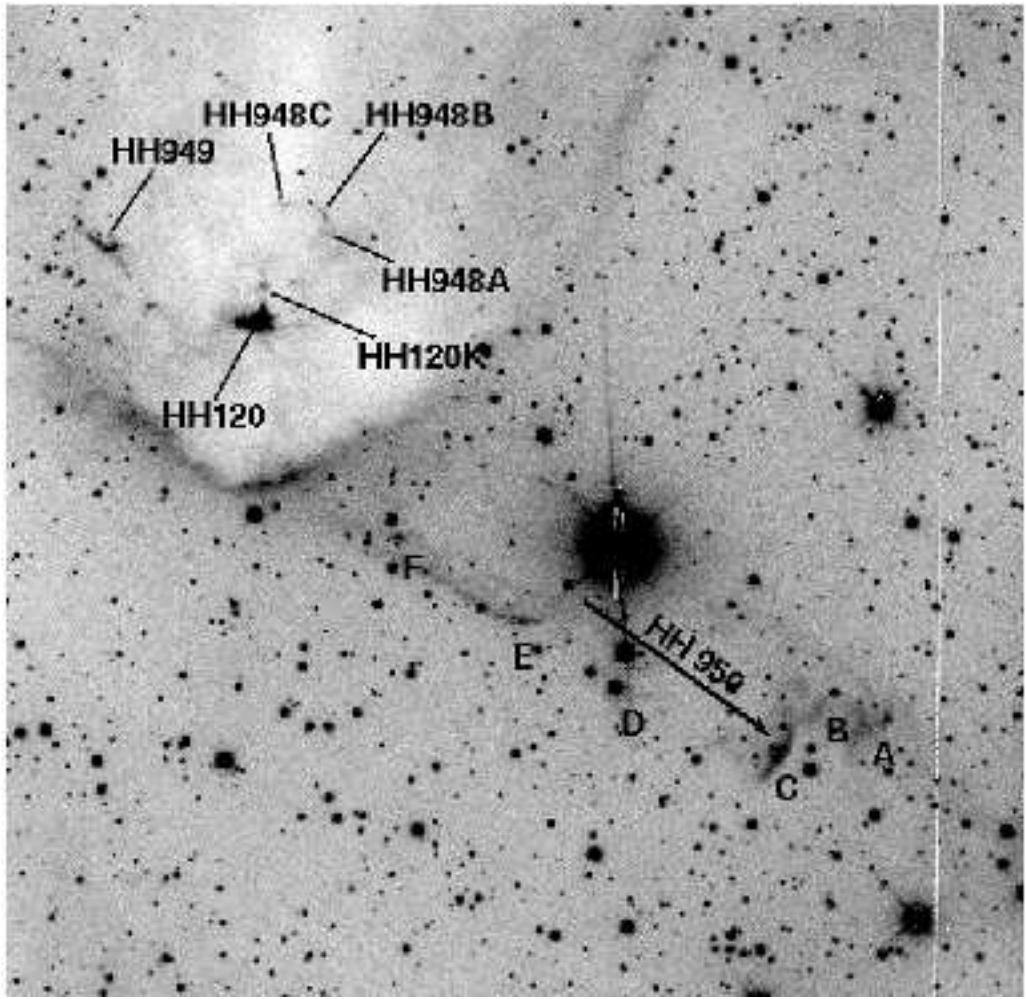}
\caption{The three NTT images have been combined into a single image
in order to have an overview of all of the HH objects in the
region. The HH knots located around HH~120 are best visible in the
[S~II] image and so are the newly discovered objetcs (HH~949, HH~948A,
B and C, HH~120K) The HH~950 knots are best resolved in the [O~II]
image (the HH~950 knots C and D only appear on this image). The field of view (FOV) of this image is 5'8$\times$5'7. North is up and east is left.}
\label{figure-combine}
\end{figure}

\begin{figure}
\centering
\includegraphics[width=1.0\textwidth]{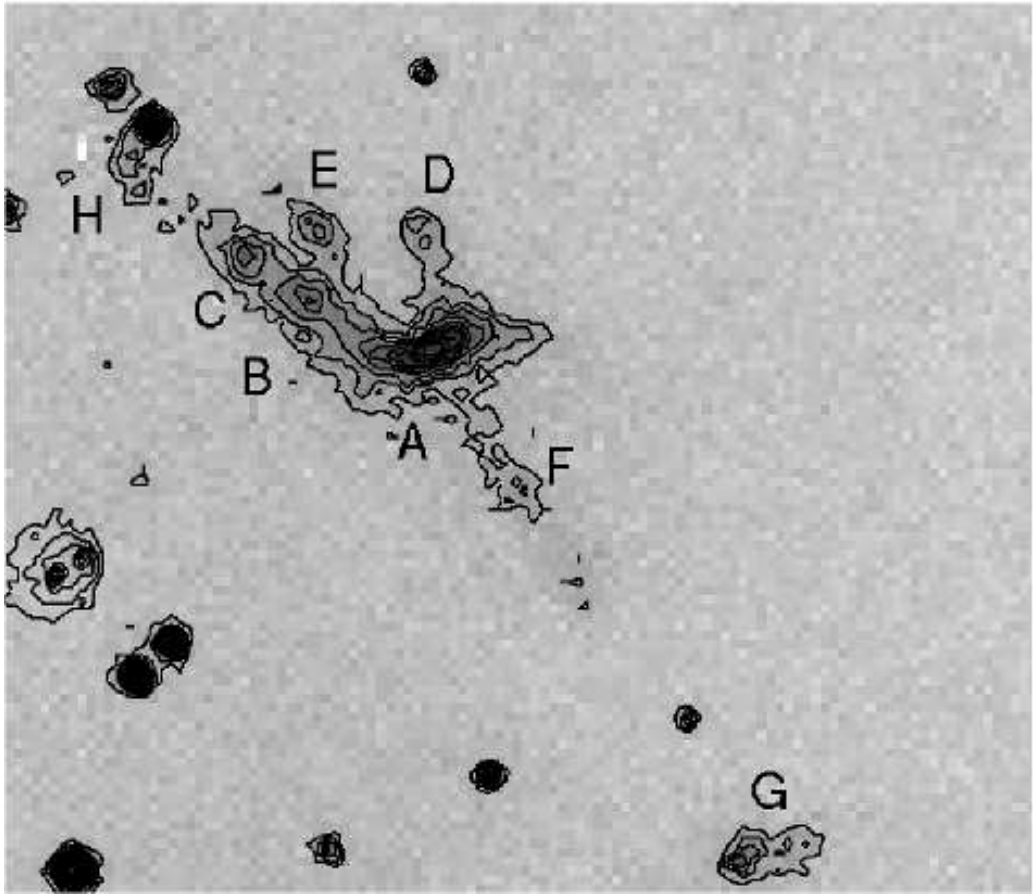}
\caption{The previous image was enlarged to show the structure of the HH~949. Also, the logarithmically scaled contours were superposed in order to better resolve the individual knots of this flow. Eight knots can be resolved. Six of them (A to F) have previously been labeled by Hodapp \& Ladd (1995) as IR knot 7, one of them (H) as IR knot 8 and G has been known as IR knot 5. The angular size of this image is 42''$\times$34''.}
\label{hh949}
\end{figure}

\begin{figure}
\centering
\includegraphics[width=1.0\textwidth]{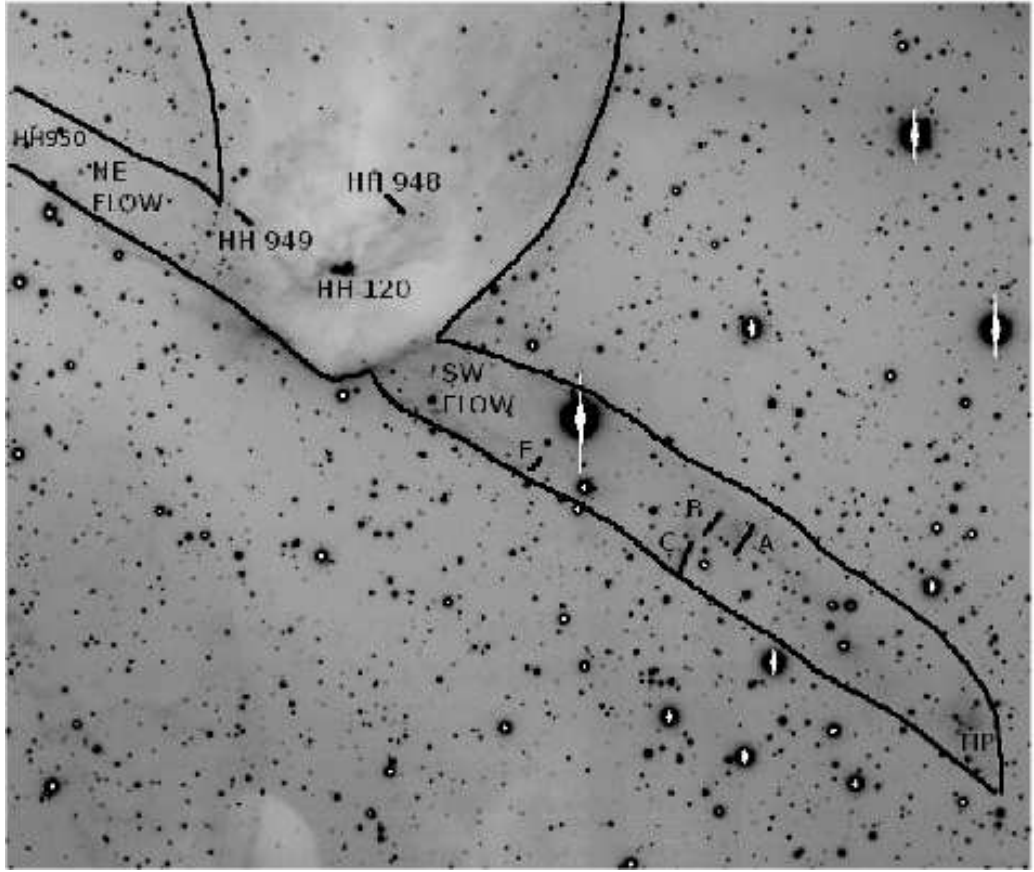}
\caption{This image includes a hand made sketch in order to make it
easier for the reader to recognize different features in the
image. The shape of the HH~950 jet and all the HH objects are
marked. The FOV of the image is 8'9$\times$7'6. North is up and east is left.}
\label{fig-draw}
\end{figure}

\begin{figure}
\centering
\includegraphics[width=1.0\textwidth]{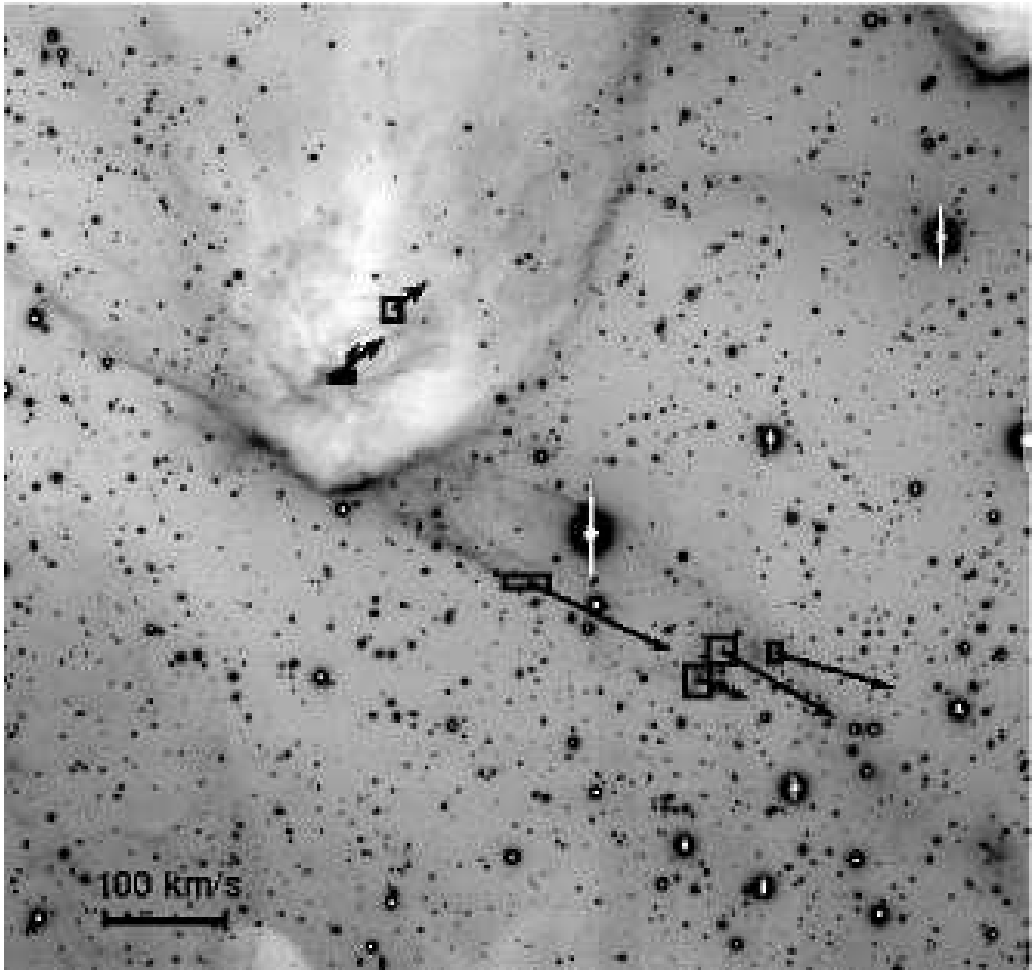}
\caption{Proper motion vectors for individual features in the CG~30
complex. The boxes indicate regions used for cross correlation. The
values for the vectors are given in Table~\ref{tab-pm}. The FOV is 8'6$\times$9'1. North is up and east is left.}
\label{fig-pm}
\end{figure}

\begin{figure}
\centering
\includegraphics[width=1.0\textwidth]{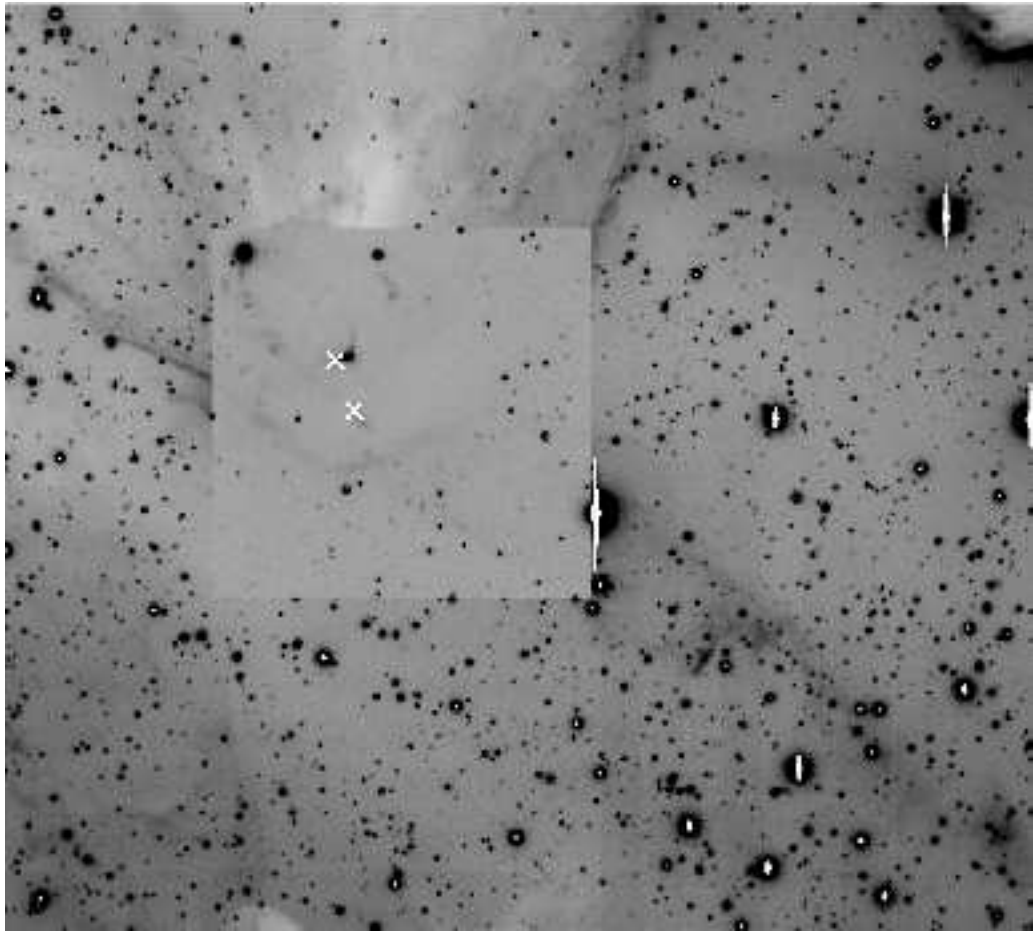}
\caption{The infrared image from Hodapp \& Ladd (1995) is superposed
onto the large Subaru image, so we can directly see the positions of
the eight infrared shocked objects and put them in the context of our
optical data. The two crosses mark the positions of the two submm
sources found by Launhardt et~al. (2000). For a detailed discussion,
see \S~4.The angular size of the image is 8'6$\times$7'7.}
\label{fig-krizi}
\end{figure}

\end{document}